\documentclass{INTERSPEECH2023}
\usepackage{comment}

\usepackage{xcolor}

\usepackage{makecell}

\title{Cascaded encoders for fine-tuning ASR models on overlapped speech}
\name{Richard Rose, Oscar Chang, and Olivier Siohan}
\address{Google Inc., New York}
\email{\small \tt \{rickrose, oscarchang, siohan\}@google.com}

\usepackage{multirow}
\begin{document}
\ninept
\maketitle
\begin{abstract}
Multi-talker automatic speech recognition (MT-ASR) has been shown to improve ASR performance
on speech containing overlapping utterances from more than one speaker.
While MT-ASR models have typically
been trained from scratch using simulated overlapping
speech datasets, there is generally an underlying goal that these
models also obtain state of the art performance on single speaker
utterances as well.
This implies that they must be competitive
with the best available fine-tuned speech models that have been
trained using massive datasets collected from a wide variety of task domains.
This paper presents an MT-ASR model formed by combining a well-trained foundation model with a multi-talker mask model in a cascaded RNN-T encoder
configuration. 
Experimental results show that the cascade configuration provides improved WER on
overlapping speech utterances with respect to a baseline multi-talker model with minimal impact on the performance achievable by the foundation
model on non-overlapping utterances.

\end{abstract}

\section{Introduction}
\label{sec-intro}

It is well known that overlapping speech exists in
utterances arising from human-human
interaction~\cite{Cetin2006,Tripathi2020}.
A study of utterances in a meetings domain found that a wide range
of behaviors relating to overlapping speech are present~\cite{Cetin2006}.
In a study of interactions in a call center domain,
roughly 12\% of the word occurrences in client–operator interactions
were found to correspond to overlapping speech~\cite{Tripathi2020}.
There has been a great deal of recent work
on multi-talker automatic speech recognition
(MT-ASR)~\cite{DongYu-2017,kanda2020,Tripathi2020,Chang-2020,lu-2021,rose2021,rose2022,sklyar22},
which attempts to improve speech recognition
from multiple overlapping speakers by decoding transcriptions from each speaker.

This paper focuses on a set of multi-talker approaches that augment the traditional single label
audio / audio-visual encoder with a mask encoder~\cite{Tripathi2020,rose2021,rose2022}.
Training involves a two pass procedure where an overlapping speech
utterance is aligned with each of multiple transcriptions
associated with the overlapping speakers.
Most of the E2E multi-talker techniques have been applied
to the case where there are two overlapping speakers; however,
there have been more recent efforts to generalize to more widely varying
overlapping speech scenarios~\cite{Raj-2021,kanda2020,Sklyar-2021,sklyar22,VonNeumann-2021}. 
All of these models have been trained from scratch using
simulated or actual overlapping speech datasets.
However, in most scenarios, there is an underlying goal that
these models obtain state of the art
performance on single speaker utterances as well
as overlapping speech utterances.
This implies that they must be competitive with the best
available fine-tuned single speaker models on single speaker utterances.

Foundation models (FMs) for ASR are large models trained on a
broad range of data sources at scale that can be applied to a
wide range of tasks~\cite{Zhang2023,Radford2023}.
They have in practice demonstrated strong generalization
and knowledge transfer capabilities.
There are many examples of training these models
in self-supervised~\cite{baevski-2020} and supervised modes~\cite{li-2021}.    
In supervised training, multiple tasks are unified by training of the model
on labeled data from these tasks~\cite{li-2021}.
Existing work has mainly focused on using supervised in-domain data to fine-tune FMs for target tasks~\cite{Chan-2021}.
Techniques such as residual adapters~\cite{Hwang-2022} have been applied to efficiently 
adapting FMs to a given target task. 

Given the scale of FMs models and their ability to generalize well across a variety of generally single speaker domains,
it makes sense to consider scenarios where a FM,
or any well trained single speaker model,
is augmented and fine-tuned to have a multi-talker capability. 
Towards this end, there are two major contributions made
by the work described in this paper.
The first is an approach for augmenting and fine-tuning a well trained
single label encoder RNN-T ASR model to perform MT-ASR decoding on
overlapping utterances.
This is done by combining an audio encoder trained from a large dataset of
single speaker utterances with a
multi-talker mask encoder in a cascaded RNN-T encoder configuration.
The second contribution is a mechanism for detecting overlapping speech
through the use of a frame-based multi-talker speech activity detector (MT-SAD).
It will be shown that a multi-talker model that decodes text
from overlapped speech also contains information about which
of multiple speakers is speaking at a given time.

Cascaded encoder configurations in E2E RNN-T models have been used as an effective approach for
unifying models that perform different tasks~\cite{narayanan-2021,chang-2022}.
For example, combining streaming and non-streaming audio encoders
in a cascade configuration was found to enhance the performance of
the non-streaming model without sacrificing the performance of the streaming model~\cite{narayanan-2021}.
In another example, a cascade combination of audio-only and audio-visual encoders
was found to improve performance on audio-visual utterances
without sacrificing performance on the audio-only task~\cite{chang-2022}.
Section~\ref{sec-system_description} of this paper introduces an approach for combining a large pre-trained audio encoder and multi-talker mask encoder in a cascade configuration.
It will be shown in Section~\ref{sec-results} that the resulting model
improves the performance of multi-talker ASR on overlapped utterances
with minimal impact on the performance of the more well trained audio encoder on single speaker utterances. 

Even when there is a significant amount of overlapping speech in the input utterances,
there is still an expectation that many utterances that are input to an MT-ASR system will be single speaker utterances.
Decoding these utterances with a two-pass MT-ASR decoder is both less
efficient and likely to provide higher WER than a SoTA single channel ASR decoder.
It is inefficient because there are multiple decoding passes, one for each expected
overlapping utterance, instead of a single decoding pass.
The WER is likely to be higher,
at least for the mask-based approach described in Section~\ref{sec-system_description},
because of the possibility
of erroneously decoding text for multiple speakers even when
speech from only a single speaker is present.
In order to reduce the errors made in multi-talker decoding on single speaker utterances,
this paper presents a mechanism for detecting overlapping speech.   
A multi-talker speech activity detector (MT-SAD) is described that detects frame based
speech activity from each of multiple overlapping speakers. 
The MT-SAD system described in Section~\ref{sec-MTSAD} detects the occurrence of multiple overlapping speakers so that MT-ASR decoding is performed only when overlapped speech is detected in an utterance.

\section{Cascaded encoder multi-talker models}
\label{sec-system_description}
This section introduces the cascaded encoder approach
to multi-talker (MT) modeling. 
First, the audio-only mask-based MT models presented in~\cite{Tripathi2020, rose2021} and the cascaded
encoder model configuration presented in~\cite{narayanan-2021,chang-2022} are described.
Second, the cascaded encoder implementation of the mask-based multi-talker model is motivated and described. 	
\subsection{MT-Baseline: Mask based multi-talker model}
\label{subsec-AV-multitalker}

\begin{figure}[htbp]
	\centering
	\vspace{-0.1in}
	\hspace{-.15in}
	\includegraphics[width=4.0cm]{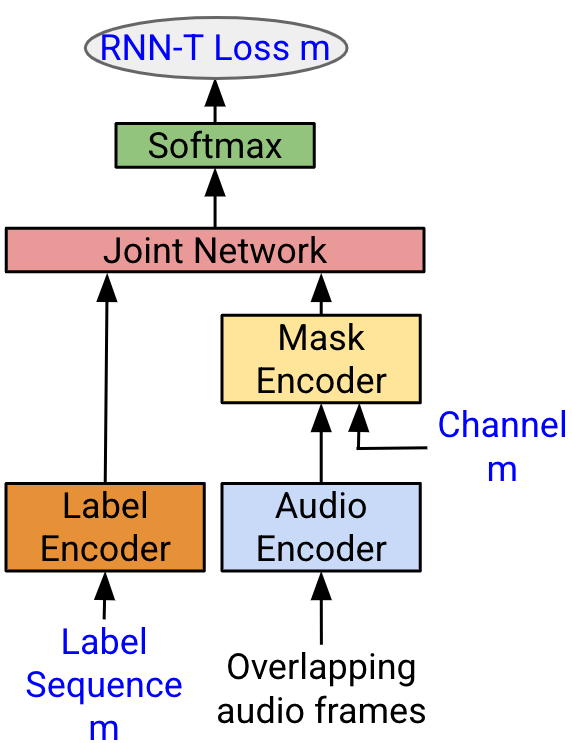}
	\vspace{-0.1in}
	\caption{MT-Baseline: Multi-talker (MT) RNN-T model.}
    \vspace{-0.1in}
	\label{fig-multitalker_review}
\end{figure}

A simplified block diagram of the audio-only MT model from~\cite{Tripathi2020} is shown in
Figure~\ref{fig-multitalker_review}.
It was shown that the single label encoder RNN-T can be
extended to the multi-talker case by adding a masking
model as shown in the figure. It is assumed in the figure that
the audio input can contain up to M overlapping utterances. In
training, it is assumed that a separate reference label sequence
exists for each of the $M$ overlapping utterances from distinct speakers. Multi-talker
training is performed by separately aligning the overlapped audio
frames to each of the $M$ label sequences.
A unique channel sequence index, $m=1,\ldots,M$, is appended to the encoded audio features
for each label sequence before the encoded audio is input to the
mask model. This serves to disambiguate speech associated
with label sequence $m$ from competing speech.
Separate RNN-T losses are computed for each of the $M$ label sequences,
and the overall RNN-T loss is the sum of channel specific RNN-T losses.
\vspace{-0.15in}
\subsection{Cascaded encoder E2E RNN-T}
\label{subsec-cascaded}

\begin{figure}[htbp]
	\centering
	\vspace{-0.1in}
	\hspace{0.0in}
	\includegraphics[width=8.5cm]{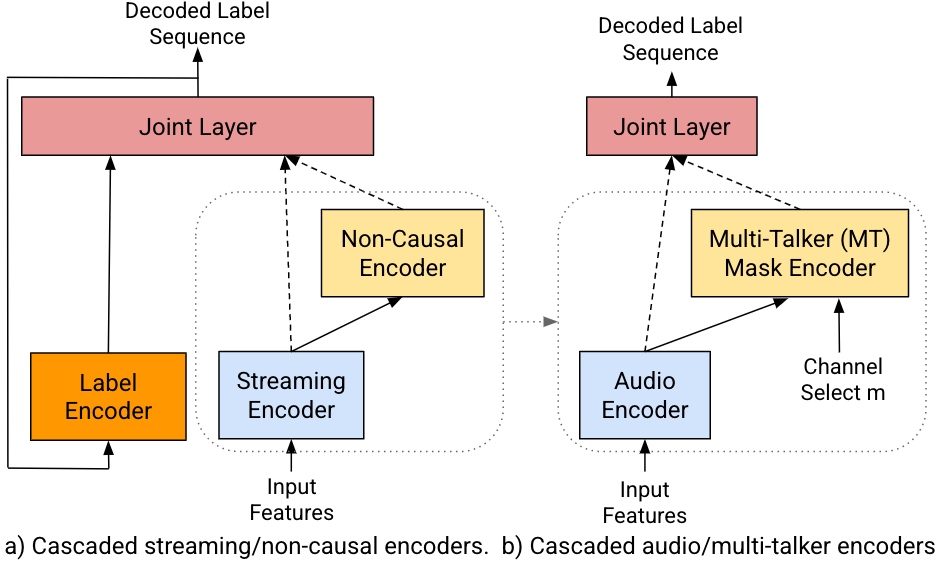}
	\vspace{-0.15in}
	\caption{a) E2E RNN-T model with cascaded streaming and non-causal encoders. 
                 b) Cascaded audio and mask encoders.}
	\label{fig-multitalker_cascade}
\end{figure}

An example of a cascaded connection of audio encoders from~\cite{narayanan-2021} is shown in Figure~\ref{fig-multitalker_cascade}a.
In this case, the audio encoder in an RNN-T model was replaced by a cascade connection of streaming
and non-causal encoders. 
The input features are first passed to a
streaming encoder, which transforms the features to a higher-level representation.
The non-causal encoder, which is connected in cascade to the causal encoder,
receives the output of the streaming encoder as input.
Both the causal and the non-causal encoders are directly connected to a shared RNN-T decoder.
The total loss is computed as the weighted sum of the RNN-T losses, $\mathcal{L}_s$
for the streaming encoder and $\mathcal{L}_n$ for the non-causal encoder,
\begin{equation}
\mathcal{L}_t = \lambda \mathcal{L}_s + (1 - \lambda) \mathcal{L}_n.
\end{equation}
This is implemented by randomly sampling in a mini-batch from the streaming / non-streaming encoder outputs 
with a sampling rate of $\lambda$.

%
\vspace{-0.2in}
\subsection{MT-Cascade: Cascaded encoder multi-talker model}
\label{subsec-cascaded-multitalker}
A simplified block diagram of the multi-talker (MT) model cascade configuration
is shown in Figure~\ref{fig-multitalker_cascade}b.
There are two advantages of this configuration over the MT-Baseline serial configuration 
of audio encoder and mask encoder shown in
Figure~\ref{fig-multitalker_review}.
First, the distribution of higher-level acoustic features can be learned for both single label and multi-label encoder RNN-T models.
The total loss for the cascade model is the sum of single label loss through the audio encoder, $\mathcal{L}_a$, and multiple label loss.
For the case of $M=2$ label sequences, this is given by
\begin{equation}
\mathcal{L}_t = \lambda \mathcal{L}_a + (1 - \lambda) (\mathcal{L}_{m1} + \mathcal{L}_{m2}), 
\end{equation}
where $\mathcal{L}_{m1}$ and $\mathcal{L}_{m2}$ correspond to the channel specific RNN-T 
losses described in Section~\ref{subsec-AV-multitalker}.
This is implemented in practice by randomly sampling overlapping utterances from a training set of overlapping and single speaker utterances with a sampling rate of $\lambda$.
A second advantage is that the mask encoder can be trained directly
on the output of an audio encoder that has been pre-trained
on data from a wide range of domains instead of being trained strictly from a smaller overlapping speech dataset.
\section{Multi-talker speech activity detection}
\label{sec-MTSAD} 
This section describes the use of a mask encoder for detecting speech activity
associated with individual speakers from overlapping utterances.
This is a reasonable goal if one considers that if the mask encoder in a multi-talker model decodes text from overlapped speech, it should also contain information
about whether or not one of multiple overlapping speakers is speaking at a given time.
The approach for multi-talker speech activity detection (MT-SAD) that is implemented here is inspired by the classifier probe
presented in~\cite{alain2017}.
In that work, an analysis of information in pre-trained networks 
is performed by inserting classifiers into intermediate
layers of a network.
The goal was to measure the level of
separability on a task that can be attained by the network features.
In~\cite{alain2017} they investigated, for example, whether there might be
information about cats in a layer of a DNN based image classifier.
The work here investigates whether there might be information about speaking activity in an E2E RNN-T multi-talker mask encoder.

\begin{figure}[htbp]
	\centering
	\vspace{-0.1in}
	\hspace{0.0in}
	\includegraphics[width=6.0cm]{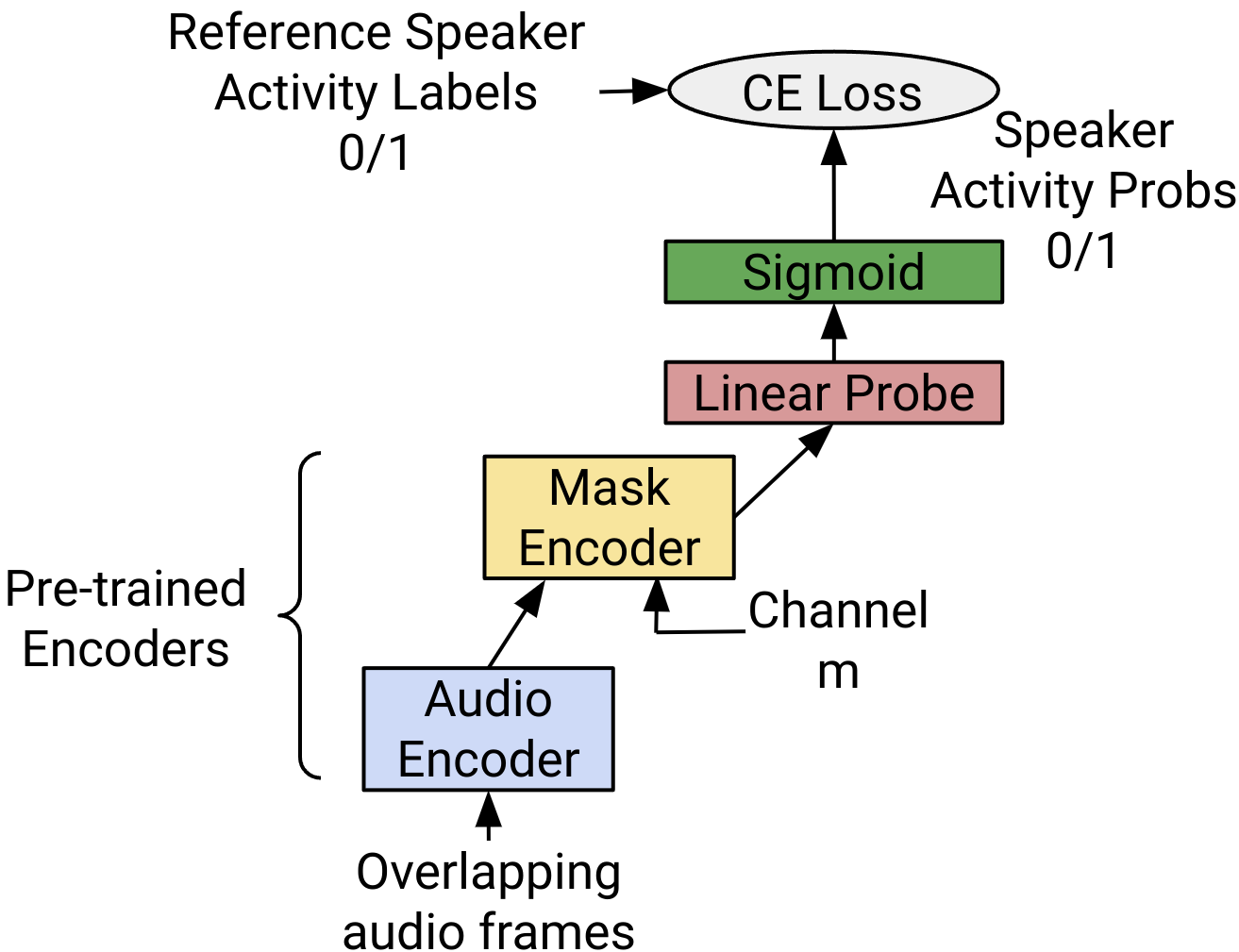}
	\vspace{0.0in}
	\caption{Speech activity detection for overlapping speech utterances.
 Audio and mask encoders are pre-trained, and remain fixed during training of probe model.}
    \vspace{-0.15in}
	\label{fig-mtsad}
\end{figure}

A block diagram of the multi-talker speech activity detector (MT-SAD) is shown in Figure~\ref{fig-mtsad}.
A single layer linear network, or ``probe'', is inserted into a layer of a pre-trained
conformer-based mask encoder. 
The inputs to the probe are the 512 dimensional output activations from the
conformer layer. 
The outputs of the probe are estimates of the probability of speaker activity
for speaker $m=1,\ldots,M$ in a given frame.
The probe is trained separate from the rest of the network with cross entropy loss
using simulated overlapping utterances. 
The frame-based reference speaker activity labels are obtained from prior knowledge of
overlap intervals in simulated overlapping speech utterances.

It will be shown in Section~\ref{sec-results} that the speaker activity
probability estimates obtained during inference from the MT-SAD model can be used
in overlapping speech detection.
This enables the scenario described in Section~\ref{sec-intro} where MT-ASR is only performed
on utterances that have been identified as containing overlapping speech and single label decoding
is performed on utterances identified as single speaker.
\vspace{-0.1in}
\section{Experimental Study}
\label{sec-study} 

This section describes an experimental study for
evaluating the performance of the
cascade configured MT-ASR models presented in
Section~\ref{sec-system_description}
and the MT-SAD system described in Section~\ref{sec-MTSAD}.
It is assumed here that there is a maximum of two overlapping speakers in the overlapped utterances.
The multi-talker experiments described in Section~\ref{sec-results} are performed using the simulated overlapping speech
training and test sets described in Section~\ref{subsec-datasets}. 
A summary of the model parameterizations is given in
Section~\ref{subsec-model}.

\subsection{Datasets}
\label{subsec-datasets}
\vspace{-.1in}
A training set of simulated two-speaker overlapped utterances was created
from a corpus of single speaker YouTube utterances.
The methodology behind the collection of the YouTube corpus can
be found in~\cite{Liao2013,shillingford2018,Makino2019,rose2021}.
The confidence island methodology in~\cite{Liao2013} facilitated the use
of user provided captions to be used as reference labels in training. 
The overlapped audio waveforms were created by taking
two of the above single speaker utterances, offsetting one in time with respect to the
other, and adding the two audio waveforms.
The emphasis in this work was to maintain accuracy for the multi-talker models on both
overlapping speech as well as single speaker utterances.

The offset used in shifting the audio signals was chosen to
provide overlap intervals randomly selected with a uniform distribution 
between 0.5 and 4.0 seconds. 
Each overlapped speech utterance was stored with two reference transcriptions and overlap interval start and end times. 
The resulting training corpus contains 15k hours of training data.
Half of the training set consists of overlapping speech utterances
and half consists of single speaker utterances.
The methodology for collecting this simulated overlapping speech dataset
is similar in some respects to the simulated overlapping speech dataset
used for the open source LibriCSS dataset~\cite{chen2020}.
However, that dataset was not used here due to fact that the
number of hours and the number of speakers used here was
several orders of magnitude larger.
A separate 150K hour non-overlapping training set,
also collected using the pipeline described
above, was used for training a ``well-trained'' single speaker ASR model. 
While this training scenario does not represent the broad range of data sources
associated with foundation model training, it provides the opportunity to 
measure the ability of the cascaded encoder configuration to match the best
performance on single speaker utterances that can be obtained by a well-trained
single label model.

Overlapped and single speaker test sets were obtained from
human transcribed utterances also taken
from YouTube videos. 
The process of forming overlapped utterances
is the same as described above for the training set. 
The test sets all contain 3601 utterances with the overlapped test utterances 
ranging in length from 2.7 to 14.7 seconds,
and the single speaker test utterances ranging in length from 2.5 to 8.0 seconds.  

The work presented in this paper abides by Google's AI Principles~\cite{AIPrinciples}.
By improving the robustness of
speech recognition systems, we hope to increase the reach of ASR
technology to a larger population of users, and use it to develop 
assistive technology.
The data and models developed in this work
are restricted to a small group of researchers
working on this project and are handled in compliance with
the European Union General Data Protection Regulation~\cite{GDPR}.

\subsection{Model Parameterization}
\label{subsec-model}
\vspace{-.1in}
Both the audio and mask encoders in all systems are conformer
models~\cite{Gulati-2020}.  
The input audio features are derived from 80 dimensional mel-warped log filter-bank energies
updated at 10 millisecond intervals.
These are concatenated to form 240 dimensional stacked input vectors with a frame rate of 33.3 frames per second.
The audio encoder consists of 17 conformer layers with internal model dimension of 512.
The mask encoder consists of 8 layers with model dimension of 512.
A cosine learning schedule was used with a 30K step warm-up and initial learning rate
of $5 \times 10^{-4}$.
The label encoders for all models use a two-layer bidirectional LSTM
with hidden dimension of 2048.
A one-layer MLP with hidden dimension of 640 is used for the joint network.

For the MT-Baseline model described in Section~\ref{subsec-AV-multitalker} and the
MT-Cascade-Scratch model, all model parameters are trained simultaneously from scratch
and no parameters are pre-trained.
For the MT-Cascade-Pretrained and the MT-Conditioned models described in Sections~\ref{subsec-cascaded-multitalker} and~\ref{sec-MTSAD} respectively, the audio encoder is pre-trained from the 150 thousand hour training set described in
Section~\ref{subsec-datasets}.

\vspace{-.1in}
\section{Experimental Results}
\label{sec-results}
This section gives the experimental results for single-talker (ST)
and MT-ASR models that are trained and evaluated on the
datasets described in Section~\ref{sec-study}.
Table~\ref{tab-st-mt-comparison} provides an illustration of the issues
introduced in Section~\ref{sec-intro} that the paper is attempting to resolve.
It displays the performance computed for ST and MT systems on
the overlapped (Overlap) and single speaker () test sets.
First, WERs are compared for ST models trained on the 150K hour and 15K hour
datasets (SingleTalker-150K and SingleTalker-15K, respectively).
The smaller training set in this case results in a 5 percent increase
in WER on single speaker utterances.
Comparing WERs in Table~\ref{tab-st-mt-comparison} on the Overlap
test set, it is clear that there is a large increase in WER when 
either ST models are evaluated on overlapping speech.
This behavior for ST decoding on overlapped speech is consistent with results observed elsewhere~\cite{Tripathi2020,kanda2020,rose2021}. 
The MT-Baseline system obtains a far lower WER on the Overlap set than the
ST systems.
However, the WER obtained for MT-Baseline on SingleSpkr is over 20 percent
higher than that obtained by SingleTalker-150K.
\begin{table}[htbp]
	\centering
        \vspace{-0.1in}
	\caption{WERs for single-talker and MT models
                on single speaker (SingleSpkr) and overlapped (Overlap) test sets.}
	\label{tab-st-mt-comparison}	
        \vspace{-0.1in}
	\begin{tabular}{|c|c|c|}
	   \hline
	  \multicolumn{3}{|c|}{\bf WER for SingleTalker and MultiTalker Models} \\\hline
		\multirow{2}{*}{\bf Model}  & \multicolumn{2}{c|}{\bf Test Sets} \\
                                        & {\bf SingleSpkr} & {\bf Overlap} \\ \hline\hline
 		SingleTalker-150K & 16.4 &  48.1 \\ \hline
            SingleTalker-15K & 17.2 &  54.1 \\ \hline
 		 MT-Baseline & 20.4 & 23.1 \\ \hline
    \end{tabular}
\vspace{-0.1in}
\end{table}

Table~\ref{tab-multitlaker-cascade-comparison}
shows the impact of the cascaded encoder implementation of the MT model.
Rows two and three show WERs for the MT cascade when all parameters are
trained from scratch
(MT-Cascade-Scratch) and the MT cascade initialized with the pre-trained audio encoder
 from SingleTalker-150K (MT-Cascade-Pretrained).
MT-Cascade-Scratch shows a small improvement in WER on Overlap compared to 
MT-Baseline, and an 11 percent WER reduction on SingleSpkr. 
MT-Cascade-Pretrained shows a 9 percent WER reduction on Overlap, and 
almost identical WER compared to the best ST model, SingleTalker-150K, on 
the SingleSpkr set.
\begin{table}[htbp]
	\centering
        \vspace{-0.1in}
	\caption{WERs for MT-Baseline and MT-Cascade models}
        \vspace{-0.1in}
	\label{tab-multitlaker-cascade-comparison}	
 
	\begin{tabular}{|c|c|c|c|}
	   \hline
	  \multicolumn{4}{|c|}{\bf WER for MultiTalker Models} \\\hline
		\multirow{2}{*}{\bf Model}  & \multicolumn{3}{c|}{\bf Test Sets} \\
                                        & {\bf SingleSpkr} & {\bf Overlap} & {\bf Ave.}\\ \hline\hline
 		 MT-Baseline & 20.4 & 23.1 & 21.7 \\ \hline
		 MT-Cascade-Scratch  & 17.9    &  22.6 & 20.2 \\ \hline
         MT-Cascade-Pretrained & 16.5 &  21.3 & 18.9 \\ \hline 
         MT-Cascade-Conditioned &  -- & -- & 19.0 \\ \hline
    \end{tabular}
\vspace{-0.1in}
\end{table}
%


The accuracy of the frame based speech activity scores generated by the MT-SAD system described in Section~\ref{sec-MTSAD} was also evaluated.
Inference is run twice through the  MT-SAD, once for each setting of the
channel select index in Figure~\ref{fig-mtsad} to generate frame-based speaker activity estimates for each of the two decoding channels.
As a result, there are separate SAD labels for each 
of the possibly overlapping speakers in an utterance.
The average frame classification accuracy
on the combined Overlap and SingleSpkr test sets was 91 percent. 
While this task is far less demanding than other speech activity
detection tasks~\cite{Joglekar-2020}, it is important to note that this performance is
obtained in the context of heavily overlapped speech.

A scenario was proposed in Section~\ref{sec-MTSAD} where
multi-talker decoding was conditioned on the probability of there being
overlapping speech in the utterance.
In this scenario, MT-SAD produces estimates of the probability that there is overlapping speech in an utterance.
These estimates are derived from the frequency of co-occurrence of frames where active speech is found by the MT-SAD for both of $M=2$ channels.
Generating these overlapping speech probability estimates during recognition does not require decoding. It simply requires
inference through the MT probe shown in Figure~\ref{fig-mtsad}.
So it is a relatively efficient means for determining
whether there is a need to perform MT-ASR decoding rather than single-talker
decoding.
\begin{figure}[htbp]
	\centering
  \vspace{-0.15in}
        \includegraphics[width=7.0cm]{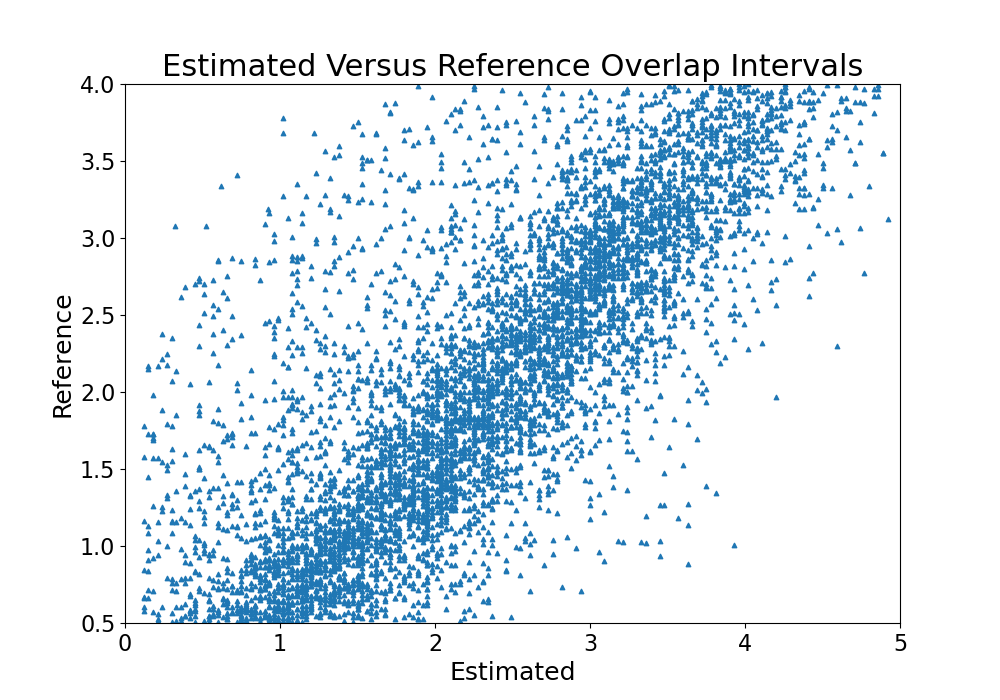}
         \vspace{-0.1in}
	\caption{Scatter plot showing the correlation between estimated and actual speaker overlap.}
 \vspace{-0.15in}
	\label{fig-scatter}
\end{figure}

Figure~\ref{fig-scatter} shows a scatter plot comparing the estimated length of
speaker overlap intervals, shown along the horizontal axis,
with prior knowledge of speaker overlap shown along the vertical axis.
It provides anecdotal evidence showing that the overlap estimates are a reasonably
good predictor of the actual length of speaker overlap.
The fourth row of Table~\ref{tab-multitlaker-cascade-comparison} (MT-Cascade-Conditioned) shows the average SingleSpkr/Overlap WER obtained
under the following scenario. Two-pass multi-talker decoding is performed when the estimated overlapping speech probability exceeds a threshold of 0.5 secs.\ and ST decoding is performed otherwise.
While there is a small percentage of overlapped utterances being misclassified as SingleSpkr,
it is clear that this is a potential scenario for augmenting ST decoding with
a multi-talker capability.

%
\section{Summary and Conclusions}
\label{sec-conclusions}
Two major developments are presented in this paper.
First, a cascaded RNN-T encoder approach is presented for augmenting
a well trained single-talker RNN-T ASR model to perform MT-ASR decoding on
overlapping utterances.
It was shown that the cascade configuration resulted in a
10 percent reduction in WER on overlapping speech
and negligible increase in WER on single
speaker utterances relative to a single-talker ASR system trained
from an order of magnitude more data.
Second, an approach for efficient detection of frame-based speaker activity from
overlapping speech utterances is presented.   
This facilitated the implementation of a ``multi-talker-conditioned''
decoding scenario that performed MT decoding only when overlapping speech was likely,
and otherwise relied on a more efficient single-talker decoder. 

\section{Acknowledgments}
The authors would like to thank Avner May and Hank Liao for work on evaluation tools,
Basi Garcia for help with data generation,
and Otavio Braga for work on input generator for RNN-T.
\vspace{-0.1in}


\ninept
\bibliographystyle{IEEEbib}
\bibliography{refs}

\begin{thebibliography}{10}

\bibitem{Cetin2006}
{\"O}.~Çetin and E.~Shriberg,
\newblock ``Analysis of overlaps in meetings by dialog factors, hot spots,
  speakers, and collection site: insights for automatic speech recognition,''
\newblock in {\em InterSpeech}, 2006.

\bibitem{Tripathi2020}
Anshuman Tripathi, Han Lu, and Hasim Sak,
\newblock ``End-to-end multi-talker overlapping speech recognition,''
\newblock in {\em ICASSP 2020}, pp. 6129--6133.

\bibitem{DongYu-2017}
Dong Yu, Morten Kolbæk, Zheng-Hua Tan, and Jesper Jensen,
\newblock ``Permutation invariant training of deep models for
  speaker-independent multi-talker speech separation,''
\newblock in {\em ICASSP 2017}, pp. 241--245.

\bibitem{kanda2020}
Naoyuki Kanda, Xuankai Chang, Yashesh Gaur, Xiaofei Wang, Zhong Meng, Zhuo
  Chen, and Takuya Yoshioka,
\newblock ``Investigation of end-to-end speaker-attributed {ASR} for continuous
  multi-talker recordings,''
\newblock in {\em IEEE Spoken Language Technology Workshop (SLT)}, 2021, pp.
  809--816.

\bibitem{Chang-2020}
{X}. Chang, {W}. Zhang, {Y}. Qian, {J}.~Le Roux, and {S}. Watanabe,
\newblock ``End-to-end multi-speaker speech recognition with transformer,''
\newblock in {\em ICASSP 2020}.

\bibitem{lu-2021}
Liang Lu, Naoyuki Kanda, Jinyu Li, and Yifan Gong,
\newblock ``Streaming end-to-end multi-talker speech recognition,''
\newblock {\em IEEE Signal Processing Letters}, April 2021.

\bibitem{rose2021}
Richard Rose, Olivier Siohan, Anshuman Tripathi, and Otavio Braga,
\newblock ``End-to-end audio-visual speech recognition for overlapping
  speech,''
\newblock in {\em InterSpeech}, 2021.

\bibitem{rose2022}
Richard Rose and Olivier Siohan,
\newblock ``End-to-end multi-talker audio-visual {ASR} using an active speaker
  attention module,''
\newblock in {\em InterSpeech}, 2022.

\bibitem{sklyar22}
Ilya Sklyar, Anna Piunova, and Christian Osendorfer,
\newblock ``{Separator-Transducer-Segmenter: Streaming Recognition and
  Segmentation of Multi-party Speech},''
\newblock in {\em Interspeech 2022}, pp. 4451--4455.

\bibitem{Raj-2021}
Desh Raj, Liang Lu, Zhuo Chen, Yashesh Gaur, and Jinyu Li,
\newblock ``Continuous streaming multi-talker {ASR} with dual-path
  transducers,''
\newblock {\em ArXiv}, vol. abs/2109.08555, 2021.

\bibitem{Sklyar-2021}
Ilya Sklyar, Anna Piunova, Xianrui Zheng, and Yulan Liu,
\newblock ``Multi-turn {RNN-T} for streaming recognition of multi-party
  speech,''
\newblock {\em ArXiv}, vol. abs/2112.10200, 2021.

\bibitem{VonNeumann-2021}
Thilo von Neumann, Keisuke Kinoshita, Christoph Boeddeker, Marc Delcroix, and
  Reinhold Haeb-Umbach,
\newblock ``Graph-{PIT}: Generalized permutation invariant training for
  continuous separation of arbitrary numbers of speakers,''
\newblock {\em arXiv preprint arXiv:2107.14446}, 2021.

\bibitem{Zhang2023}
Yu~Zhang et~al,
\newblock ``Google {USM}: Scaling automatic speech recognition beyond 100
  languages,''
\newblock {\em arXiv preprint arXiv:2303.01037}, 2023.

\bibitem{Radford2023}
A.~Radford, J.~W. Kim, T.~Xu, G.~Brockman, C.~McLeavey, and I.~Sutskever,
\newblock ``Robust speech recognition via large-scale weak supervision,''
\newblock {\em arXiv preprint arXiv:2303.01037}, 2023.

\bibitem{baevski-2020}
Alexei Baevski, Yuhao Zhou, Abdelrahman Mohamed, and Michael Auli,
\newblock ``wav2vec 2.0: {A} framework for self-supervised learning of speech
  representations,''
\newblock in {\em NeurIPS 2020}, December, 2020.

\bibitem{li-2021}
Bo~Li, Ruoming Pang, Tara~N. Sainath, Anmol Gulati, Yu~Zhang, James Qin, Parisa
  Haghani, W.~Ronny Huang, Min Ma, and Junwen Bai,
\newblock ``Scaling end-to-end models for large-scale multilingual {ASR},''
\newblock in {\em IEEE Automatic Speech Recognition and Understanding Workshop
  (ASRU)}, 2021.

\bibitem{Chan-2021}
William Chan, Daniel~S. Park, Chris~A. Lee, Yu~Zhang, Quoc~V. Le, and Mohammad
  Norouzi,
\newblock ``Speechstew: Simply mix all available speech recognition data to
  train one large neural network,''
\newblock 2021.

\bibitem{Hwang-2022}
Dongseong Hwang, Ananya Misra, Zhouyuan Huo, Nikhil Siddhartha, Shefali Garg,
  David Qiu, Khe~Chai Sim, Trevor Strohman, Françoise Beaufays, and Yanzhang
  He,
\newblock ``Large-scale {ASR} domain adaptation using self- and semi-supervised
  learning,''
\newblock in {\em ICASSP 2022}.

\bibitem{narayanan-2021}
Arun Narayanan, Tara~N. Sainath, Ruoming Pang, Jiahui Yu, Chung-Cheng Chiu,
  Rohit Prabhavalkar, Ehsan Variani, and Trevor Strohman,
\newblock ``Cascaded encoders for unifying streaming and non-streaming {ASR},''
\newblock in {\em ICASSP 2021}, pp. 5629--5633.

\bibitem{chang-2022}
Oscar Chang, Otavio Braga, Hank Liao, Dmitriy Serdyuk, and Olivier Siohan,
\newblock ``On robustness to missing video for audiovisual speech
  recognition,''
\newblock {\em Transactions on Machine Learning Research}, 2022.

\bibitem{alain2017}
Guillaume Alain and Yoshua Bengio,
\newblock ``Understanding intermediate layers using linear classifier probes,''
\newblock in {\em ICLR 2017}.

\bibitem{Liao2013}
H.~{Liao}, E.~{McDermott}, and A.~{Senior},
\newblock ``Large scale deep neural network acoustic modeling with
  semi-supervised training data for {YouTube} video transcription,''
\newblock in {\em IEEE Workshop on Automatic Speech Recognition and
  Understanding (ASRU) 2013}, pp. 368--373.

\bibitem{shillingford2018}
Brendan Shillingford, Yannis Assael, Matthew~W. Hoffman, Thomas Paine, Cían
  Hughes, Utsav Prabhu, Hank Liao, Hasim Sak, Kanishka Rao, Lorrayne Bennett,
  Marie Mulville, Ben Coppin, Ben Laurie, Andrew Senior, and Nando de~Freitas,
\newblock ``Large-scale visual speech recognition,''
\newblock in {\em InterSpeech}, 2018.

\bibitem{Makino2019}
T.~{Makino}, H.~{Liao}, Y.~{Assael}, B.~{Shillingford}, B.~{Garcia},
  O.~{Braga}, and O.~{Siohan},
\newblock ``Recurrent neural network transducer for audio-visual speech
  recognition,''
\newblock in {\em 2019 IEEE ASRU Workshop}, pp. 905--912.

\bibitem{chen2020}
Zhuo Chen, Takuya Yoshioka, Liang Lu, Tianyan Zhou, Zhong Meng, Yi~Luo, Jian
  Wu, Xiong Xiao, and Jinyu Li,
\newblock ``Continuous speech separation: dataset and analysis,''
\newblock in {\em ICASSP 2020}.

\bibitem{AIPrinciples}
Google,
\newblock ``Artificial intelligence at {G}oogle: Our principles,''
  \url{https://ai.google/principles/}.

\bibitem{GDPR}
European~Union Law,
\newblock ``Regulation ({EU}) 2016/679 of the {E}uropean {P}arliament and of
  the {C}ouncil of 27 {A}pril 2016 on the protection of natural persons with
  regard to the processing of personal data and on the free movement of such
  data, and repealing directive 95/46/{EC} ({General Data Protection
  Regulation}),''
  \url{https://eurlex.europa.eu/legal-content/EN/TXT/?uri=CELEX}.

\bibitem{Gulati-2020}
Anmol Gulati, James Qin, Chung-Cheng Chiu, Niki Parmar, Yu~Zhang, Jiahui Yu,
  Wei Han, Shibo Wang, Zhengdong Zhang, Yonghui Wu, and Ruoming Pang,
\newblock ``Conformer: Convolution-augmented transformer for speech
  recognition,''
\newblock in {\em InterSpeech 2020}.

\bibitem{Joglekar-2020}
A.~Joglekar, J.~H. Hansen, M.~C. Shekar, and A.~Sangwan,
\newblock ``Fearless {S}teps challenge (fs-2): Supervised learning with massive
  naturalistic apollo data,''
\newblock in {\em InterSpeech 2020}.

\end{thebibliography}
\end{document}